\documentclass[%
superscriptaddress,
showpacs,
amsmath,amssymb,
prb,
preprint,
floatfix,
endfloats*
]{revtex4-1}

\usepackage{graphicx}
\usepackage{bm}
\usepackage{color}

\begin{document}

\title{Site-controlled InGaN/GaN single-photon-emitting diode}

\author{Lei Zhang}
\affiliation{Department of Physics, University of Michigan, 450 Church St., Ann Arbor, Michigan 48109, USA}
\author{Chu-Hsiang Teng}
\affiliation{Department of Electrical Engineering and Computer Science, University of Michigan, 1301 Beal Ave., Ann Arbor, Michigan 48109, USA}
\author{Pei-Cheng Ku}
\email{peicheng@umich.edu}
\affiliation{Department of Electrical Engineering and Computer Science, University of Michigan, 1301 Beal Ave., Ann Arbor, Michigan 48109, USA}
\author{Hui Deng}
\email{dengh@umich.edu}
\affiliation{Department of Physics, University of Michigan, 450 Church St., Ann Arbor, Michigan 48109, USA}

\date{\today}

\begin{abstract}
We report single-photon emission from electrically driven site-controlled InGaN/GaN quantum dots, fabricated from a planar light-emitting diode structure containing a single InGaN quantum well using a top-down approach. The location, dimension, and height of each single-photon-emitting diode are controlled lithographically, providing great flexibility for chip-scale integration.
\end{abstract}

\pacs{0000.0000}

\maketitle

\label{Introduction}
Single-photon sources (SPSs) constitute a critical resource for a wide range of applications including quantum communication \cite{Bennett1984}, optical quantum computing \cite{Pooley2012, OBrien2003}, and precision measurement \cite{Xiao1987, Polzik1992}. Quantum dots (QDs) based on III-N semiconductors have become a promising candidate as a practical on-demand and sub-Poissonian SPS. With a large exciton binding energy (28 meV in bulk GaN), III-N QDs allow single-photon emission up to room temperature \cite{Holmes2014}. Although many other materials also have the potential for high-temperature SPSs \cite{Michler2000a,Lounis2000,Bounouar2012,Morfa2012,Mizuochi2012,Castelletto2013}, III-N materials possess a few distinctive advantages. First, III-N materials are relatively mature and have been heavily developed for optoelectronics and power electronics. Second, a scalable QD fabrication technique has been recently established for III-N materials \cite{Lee2011a,Zhang2013}. Last but not the least, the emission wavelength of III-N QDs is tunable across a wide spectral range, covering both the range where the detector performance of a silicon avalanche photodiode peaks and telecommunication wavelengths that can be suitably transmitted through standard optical fibers. Toward a practical chip-scale SPS, two critical requirements -- electrical-injection and site control -- have both been demonstrated for III-N QDs \cite{Zhang2013,Deshpande2013a,Holmes2014}, but only separately. Combining these two requirements has been a challenge from a bottom-up fabrication approach or in a non-planar structure, such as micro-pyramids \cite{Edwards2004}, inversed micro-pyramids \cite{Baier2004a}, and nanowires \cite{Holmes2014}. In these structures, the QDs are typically sandwiched between non-planar wetting layer and barrier layers, thus susceptible to high serial resistance and large leaky current. In this work, using top-down site-controlled QDs, we apply standard thin-film epitaxy to form a P-i-N structure as in a typical III-N light-emitting diode (LED), where the current-pathway naturally overlaps with the QD active region. With such a structure, we demonstrate for the first time electrically driven single-photon emission from site-controlled InGaN/GaN QDs.

\begin{figure}
	\includegraphics[width = 0.45 \textwidth]{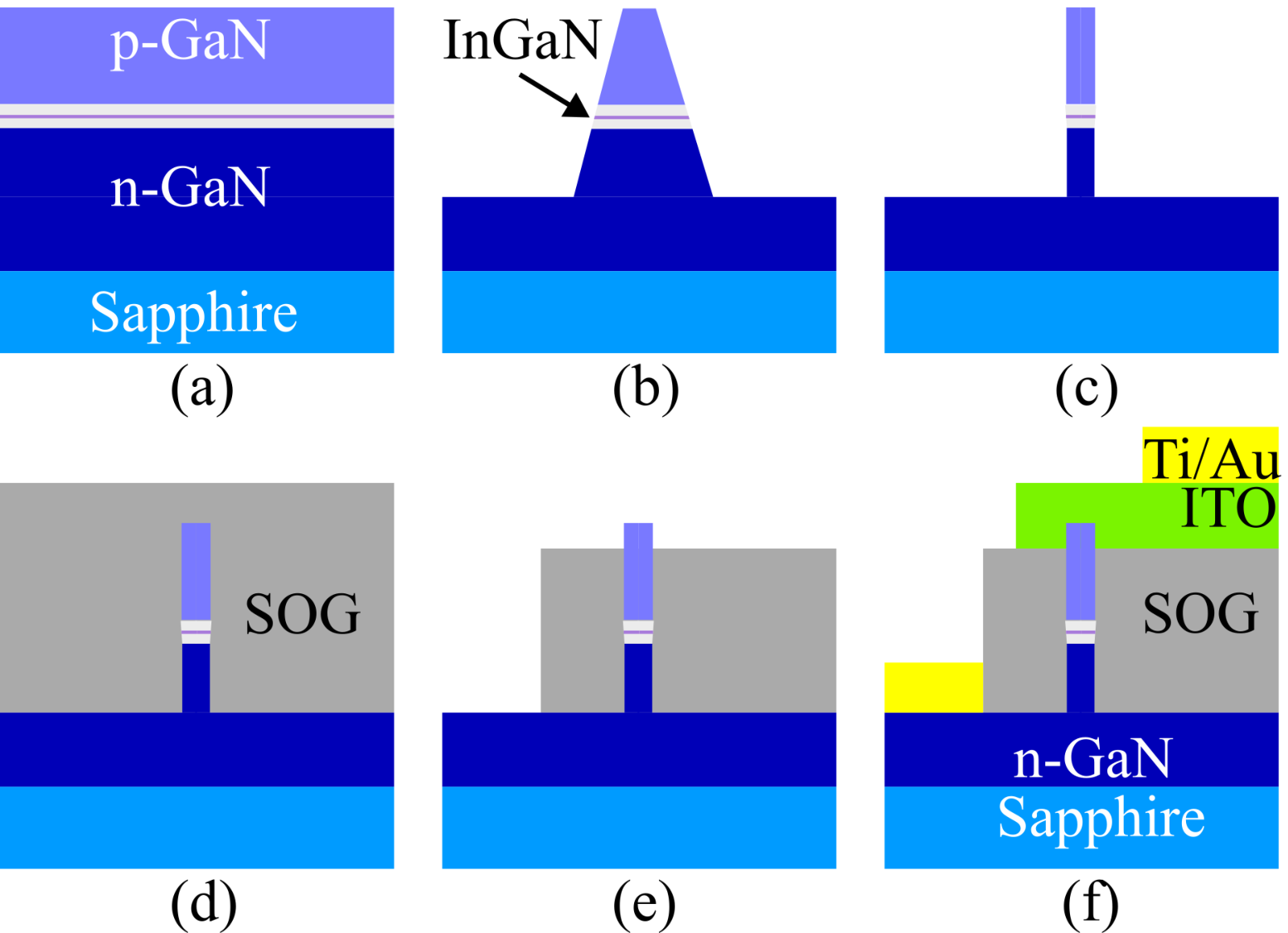}
	\caption{\label{fig:fabrication} Schematic of sample fabrication processes: (a) epitaxial growth using MOCVD on (0001) sapphire; (b) patterning of InGaN nanopillars using EBL and ICP-RIE; (c) formation of vertical nanopillar sidewalls; (d) planarization with SOG; (e) SOG etchback to expose p-GaN; (f) formation of metal contacts.}
\end{figure}

\begin{figure}
\includegraphics[width = 0.45\textwidth]{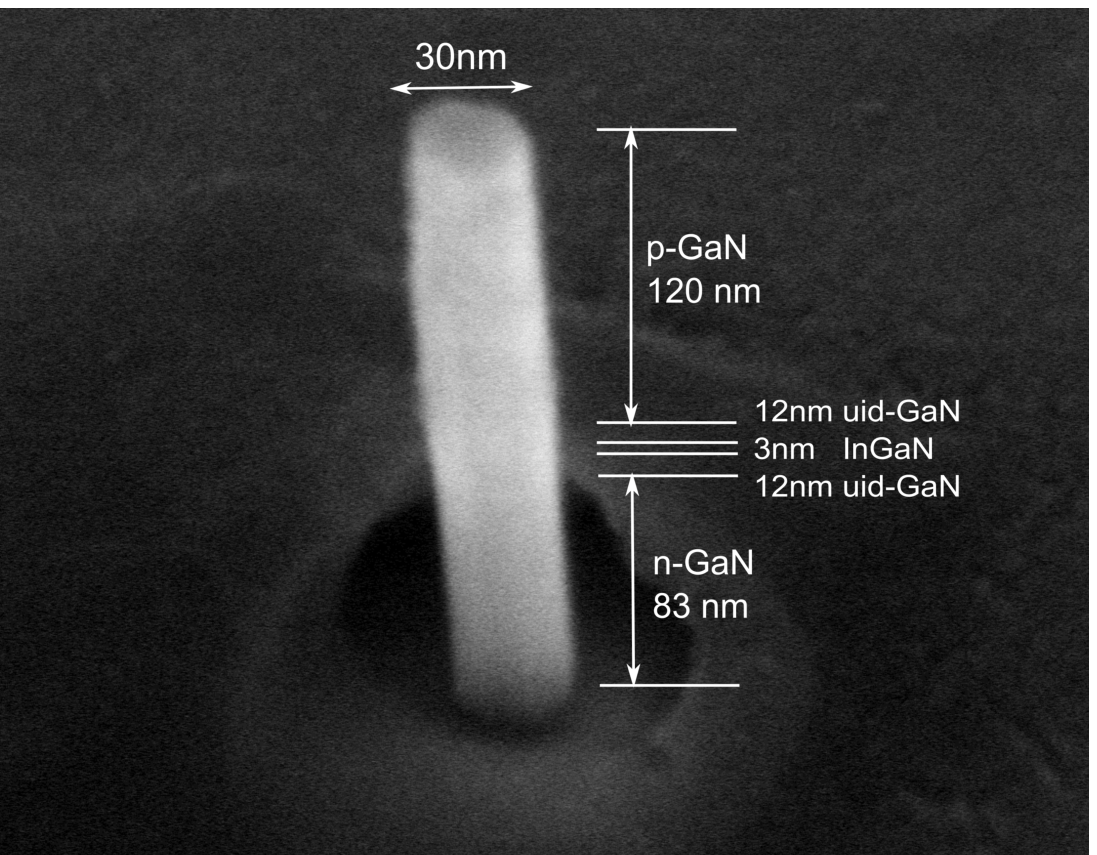}
  \caption{\label{fig:sample} 45$^\circ$-view SEM image of a nanopillar containing a single site-controlled InGaN QD in a GaN pin junction. Two 12-nm thick unintentionally doped (uid) GaN layers are used as the quantum barrier.}
\end{figure}

\label{Sample}
Figure~\ref{fig:fabrication} illustrates the fabrication process of the InGaN QD SPS. The epitaxial structure (Fig.~\ref{fig:fabrication}(a)) is similar to a standard InGaN based LED. A single InGaN/GaN quantum-well (QW) with a nominal indium composition of 15\% and 3nm thickness is sandwiched between p-GaN and n-GaN. The epitaxial growth was performed using metalorganic chemical vapor deposition (MOCVD) on a (0001)-oriented planar sapphire substrate. For simplicity, no electron nor hole blocking layer was employed but can be added in future work. After sample growth, nanopillar structures were lithographically defined (Fig.~\ref{fig:fabrication}(b)) using electron-beam lithography and inductively-coupled plasma reactive-ion etching (ICP-RIE). The details of the fabrication procedures and optical properties of the InGaN nanopillars have been reported elsewhere \cite{Lee2011a, Zhang2013, Zhang2014f, Teng2015}. The slanted nanopillar sidewall was made vertical using a 2\% buffered KOH solution (AZ400) (Fig.~\ref{fig:fabrication}(c)). The KOH etching also helps remove the plasma damaged sidewall and reduce the density of surface defect states \cite{Li2011a}. Fig.~\ref{fig:sample} shows the scanning-electron-microscope (SEM) image of a single nanopillar of 30~nm diameter and 230~nm height. Due to strain-relaxation at the sidewall, an effective confinement potential is formed for the excitons inside the QD across the QD cross section, which tightly confines the exciton at the center of the QD and shields it from the sidewall \cite{Zhang2014f}. 

After wet etching, the sample was planarized by a 580~nm thick spin-on-glass (SOG) layer followed by curing at 400~$^\circ$C for an hour (Fig.~\ref{fig:fabrication}(d)). The SOG was then etched back with CF$_4$-CHF$_3$ plasma to expose p-GaN for the p-contact (Fig.~\ref{fig:fabrication}(e)), an indium-tin-oxide (ITO) conductive film deposited by DC sputtering, patterned using HCl and annealed at 400~$^\circ$C for 5~minutes in the forming gas. Finally, both n- and p-contacts were metalized using Ti/Au (40~nm/500~nm) pads (Fig.~\ref{fig:fabrication}(f)). The room temperature carrier concentrations in p-GaN and n-GaN were estimated to be $p = 1\times 10^{17}$~cm$^{-3}$ and $n = 1\times 10^{18}$~cm$^{-3}$, respectively.

\label{Setup}

We performed micro-electroluminescence ($\mu$-EL) and micro-photoluminescence ($\mu$-PL) on individual nanopillars using the same optical setup as reported earlier \cite{Zhang2013, Zhang2014}. The sample was mounted in a temperature-stabilized continuous-flow cryostat. For EL measurements, the DC current was supplied by a source-meter unit. For PL measurements, a pulsed 370~nm laser obtained from frequency doubling a 740~nm mode-locked Ti:Sapphire laser was used as the excitation. The excitation wavelength was chosen such that only the QD and not GaN was excited. The EL and PL emission was collected by an objective lens with a numerical aperture (NA) of 0.5 and a 4~mm focal length. The excitation laser was removed using a spectral filter. The emission was then projected onto the confocal plane of a pair of identical lenses with a 75~mm focal length. A pinhole with a 25~$\mu$m diameter was placed on the confocal plane to isolate the emission from a single QD \cite{Zhang2013}. The emission spectrum was recorded by a liquid nitrogen-cooled CCD camera attached to a monochromator with 0.1~nm wavelength resolution at $\sim 400$~nm. The second-order correlation ($g^{(2)}(t)$) function was measured using a Hanbury Brown-Twiss (HBT) interferometer consisting of a beamsplitter, two avalanche photo-detectors (APDs) and a time correlator (TC). Each of the APD-TC arms had a time resolution of $\sim 200$~ps.

\label{EL}

\begin{figure}
\includegraphics[width = 0.48\textwidth]{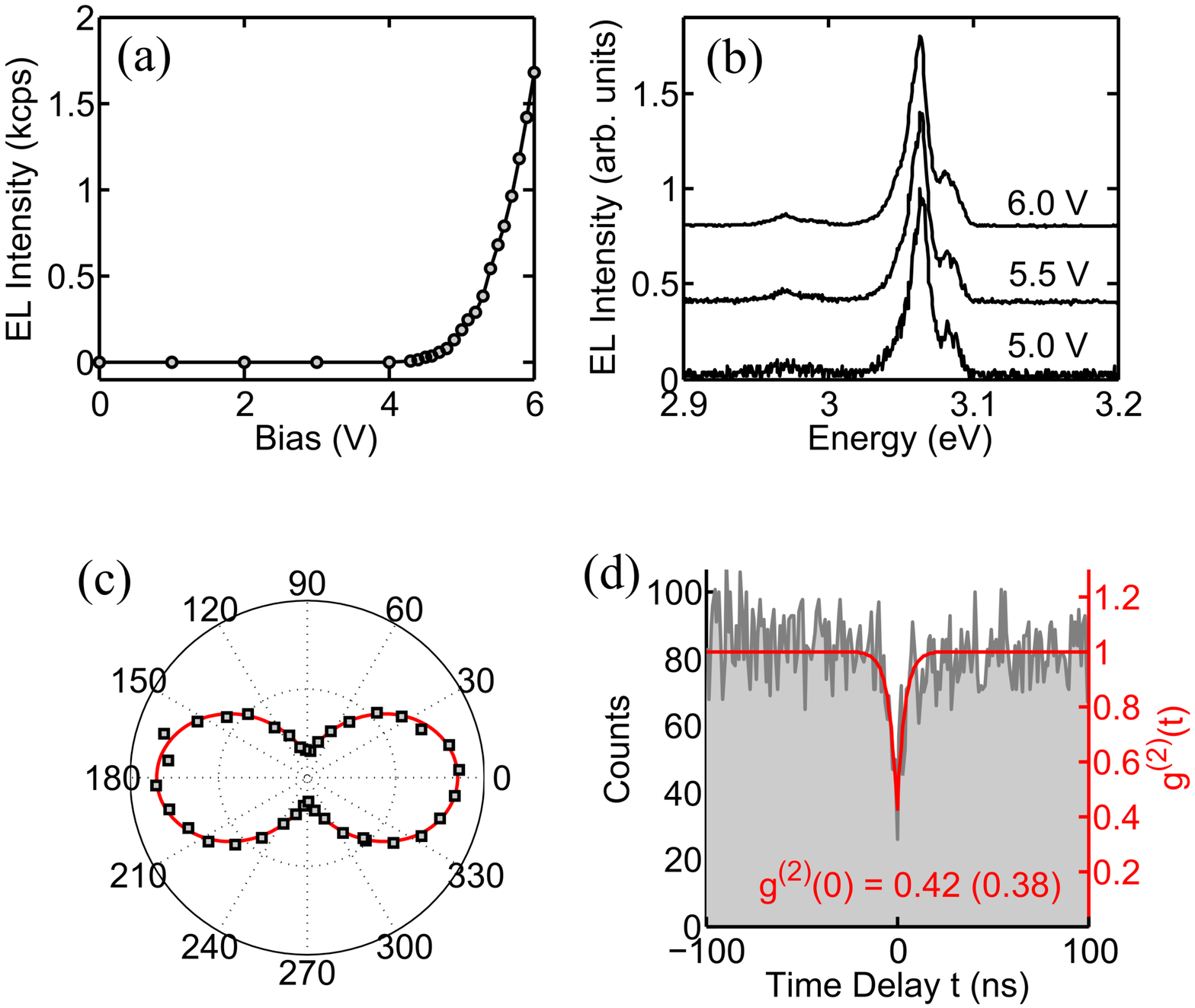}
\caption{\label{fig:el} EL properties at 10~K. (a) The integrated EL intensity vs. applied bias. (b) The EL spectra from a single QD at various bias voltages. All spectra are normalized to their maximum intensity and shifted vertically for convenience of viewing. (c) The EL intensity vs. angle of the polarization at 5.7~V forward bias. The solid line is the fitting curve. (d) The $g^{(2)}(t)$ of the EL at 5.7~V forward bias without background subtraction. The solid line is a fitting curve.}
\end{figure}

We first study the EL properties of a single nanopillar. Fig.~\ref{fig:el}(a) shows the EL intensity as a function of the applied voltage. At around 4~V forward bias, a turn-on behavior was observed, indicating the presence of both electrons and holes in the InGaN region. The EL spectra at different forward voltages above 4~V are shown in Fig.~\ref{fig:el}(b). Each spectrum is composed of a dominant zero-phonon line (ZPL) at 3.06~eV with a weak shoulder $\sim$20~meV on the higher energy side and an optical-phonon peak at 2.97~eV. The 90~meV optical-phonon energy is consistent with our previous results \cite{Zhang2013}. The ZPL and its higher energy shoulder are most likely due to emission from single exciton (X) and one of its many negatively charged states (X$^{n-}$, $n = 1$, 2, ...). This is because compared to the holes in the p-GaN region, electrons in the n-GaN region have a higher concentration and mobility ($\sim$100~cm$^2$V$^{-1}$s$^{-1}$ for electrons \cite{Gotz1996} and $\sim$10~cm$^2$V$^{-1}$s$^{-1}$ for holes\cite{Nakayama1996}). Therefore electrons arrive at the InGaN QD at a bias $V_{bias}<4$~V, before holes arrive at a higher voltage. The broad linewidth of the ZPL is attributed to the spectral diffusion as a result of local charge fluctuation caused by the current flow and the permanent dipole moment possessed by the exciton due to a strong piezoelectric field along the growth direction.

The EL is linearly polarized as shown in Fig.~\ref{fig:el}(c), which is taken by rotating a half-wave plate in front of a linear polarizer. The EL intensity vs. polarization angle ($\theta$) data are fitted using the equation $(I_1 - I_2)/(I_1+I_2)\times\cos^2(\theta + \theta_0) + (I_2)/(I_1+I_2)$. The degree of linear polarization was obtained from $(I_1 - I_2)/(I_1+I_2)=0.70 \pm 0.04$. The polarization angle $\theta_0$ is random among different QDs and does not correspond to any specific crystal orientation. Linearly polarized emission has been observed in InGaN QDs including electrically driven ones \cite{Winkelnkemper2007, Hsu2011, Zhang2013, Deshpande2013}. It has been attributed to the anisotropy in the InGaN lateral dimension \cite{Teng2015}. In our QDs, anisotropy can be caused by both the EBL and RIE processes. The anisotropic lateral shape leads to an asymmetric strain profile which mixes A and B exciton states, resulting in the linearly polarized emission \cite{Winkelnkemper2007, Teng2015}.

To demonstrate the single-photon nature of the EL, we performed second-order correlation ($g^{(2)}$) measurement to the ZPL. The result is shown in Fig.~\ref{fig:el}(d) which exhibits a strong antibunching behavior. The data can be fitted by a simple expression $g^{(2)}(t) = g^{(2)}(0) + (1-g^{(2)}(0))( 1 - exp(-|t|/\tau))$ yielding $g^{(2)}(0) = 0.42$ (or $g^{(2)}(0) = 0.38$ after adjusting for the APD dark counts \cite{Brouri2000}) and $\tau = 4$~ns. The non-zero $g^{(2)}(0)$ is due to higher order multi-exciton emission whose quantum efficiency is  largely suppressed by the non-radiative recombination at the nanopillar sidewall \cite{Zhang2013}. The parameter $\tau$ represents the decay time of the ZPL. The slow decay time is due to the strong polarization field in the InGaN region which reduces the overlap between the electron and hole wavefunctions \cite{Jarjour2007a}.

\label{IV}

\begin{figure}
\includegraphics[width = 0.4\textwidth]{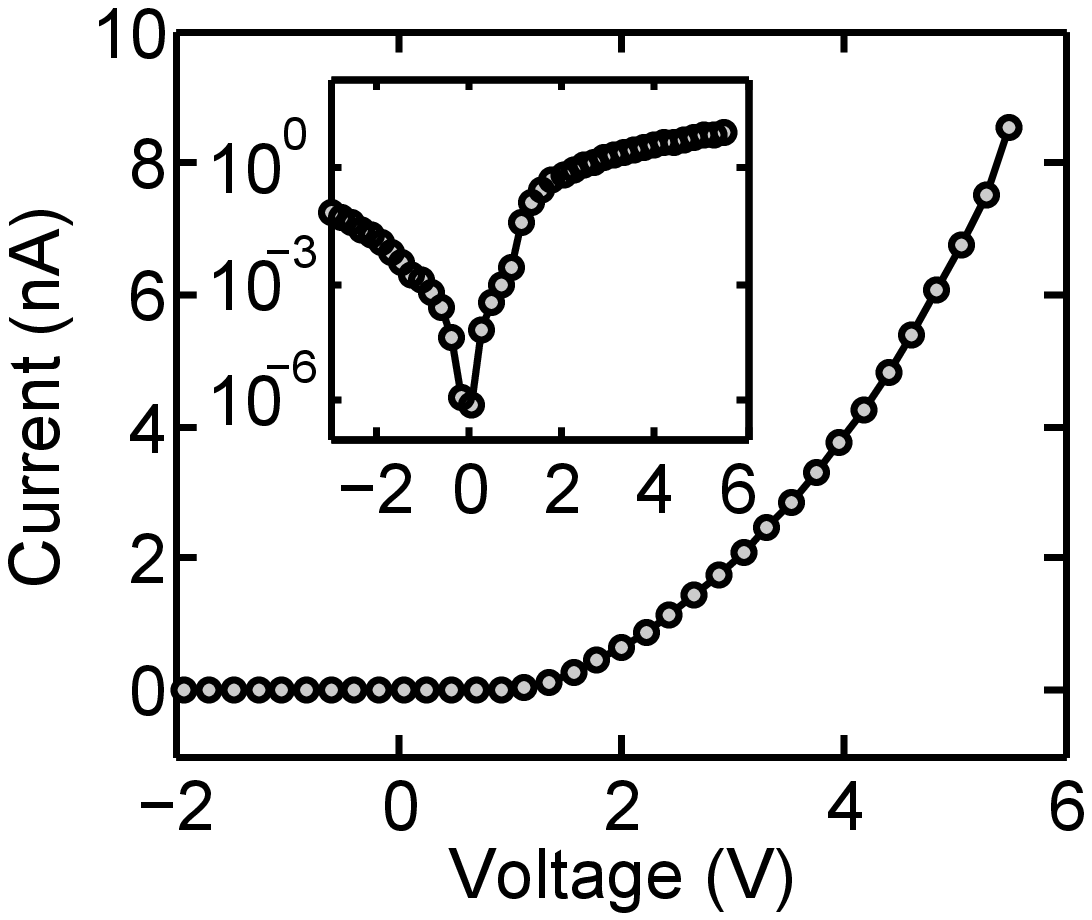}
\caption{\label{fig:iv} The current-voltage (I-V) characteristic of the DIN at 10~K. The upper-left inset is the semi-log of the I-V curve. The lower-right inset illustrates the effective circuit in which the pin-junction and the ITO/p-GaN Schottky contact are connected in series.}
\end{figure}

The current-voltage (I-V) characteristics from 8100 nanopillars connected in parallel is shown in Fig.~\ref{fig:iv}. At a small forward bias of $V_{bias}<2$~V, the current increases exponentially (inset of Fig.~\ref{fig:iv}) as the bias increases, which is expected from an ideal pin-junction. However, as $V_{bias}$ ramps above $2$~V the increase of the current becomes slow and nearly linear. This suggests the presence of a non-negligible series resistance, which is attributed to the non-ohmic ITO/p-GaN contact \cite{Margalith1999}. Using the slope of the I-V curve at $V_{bias}>2$~V and the diameter of the nanopillar, we can estimate the specific contact resistivity at the ITO/p-GaN contact to be $\sim 10^{-2}$~$\Omega$-cm$^2$, consistent with the reported values ranging from $10^{-1}$ to $10^{-3}$~$\Omega$-cm$^2$. \cite{Kim2001, Ding2012, Choi2013c}

\begin{figure}
\includegraphics[width = 0.4\textwidth]{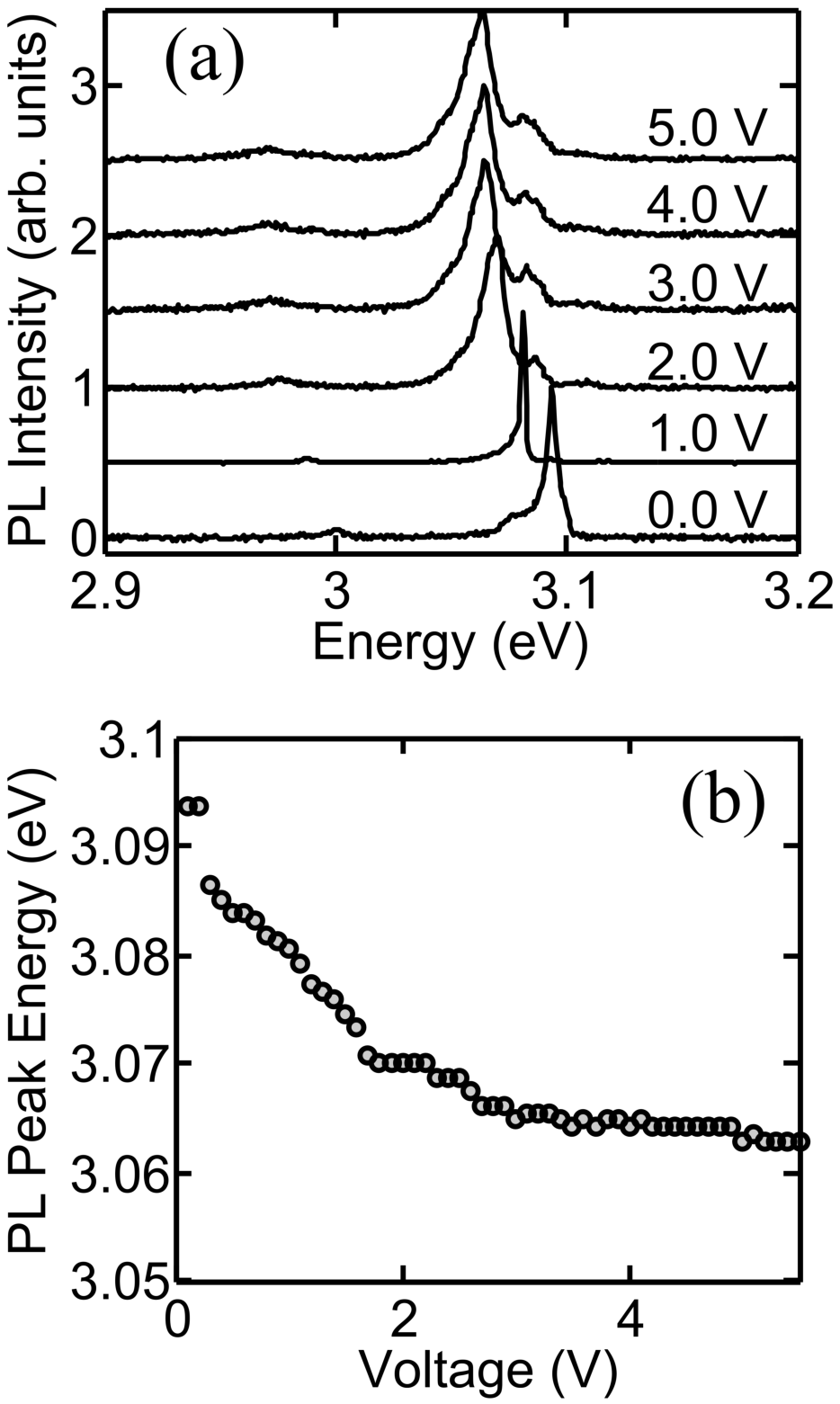}
\caption{\label{fig:pl} PL properties at 10~K. (a) The PL spectra of the DIN at several applied voltages. (b) The PL peak energy vs. the applied voltage.}
\end{figure}

\label{PL}
The above discussion of the nanopillar electrical properties was verified by the bias-dependent PL. The PL spectra at different bias voltages subject to a fixed laser excitation intensity of 100~W/cm$^2$ are shown in Fig.~\ref{fig:pl}(a). At 0V, the PL emission peaks at 3.09~eV, 30 meV higher than the EL energy in Fig.~\ref{fig:el}. As $V_{bias}$ increases to 2~V the PL peak energy redshifts at a rate of $\sim -10$~meV/V. As $V_{bias}$ increases above 2~V, the redshift stops at 3.06~eV. This trend is summarized in Fig.~\ref{fig:pl}(b) with a finer bias increment.

\begin{figure}
\includegraphics[width = 0.4\textwidth]{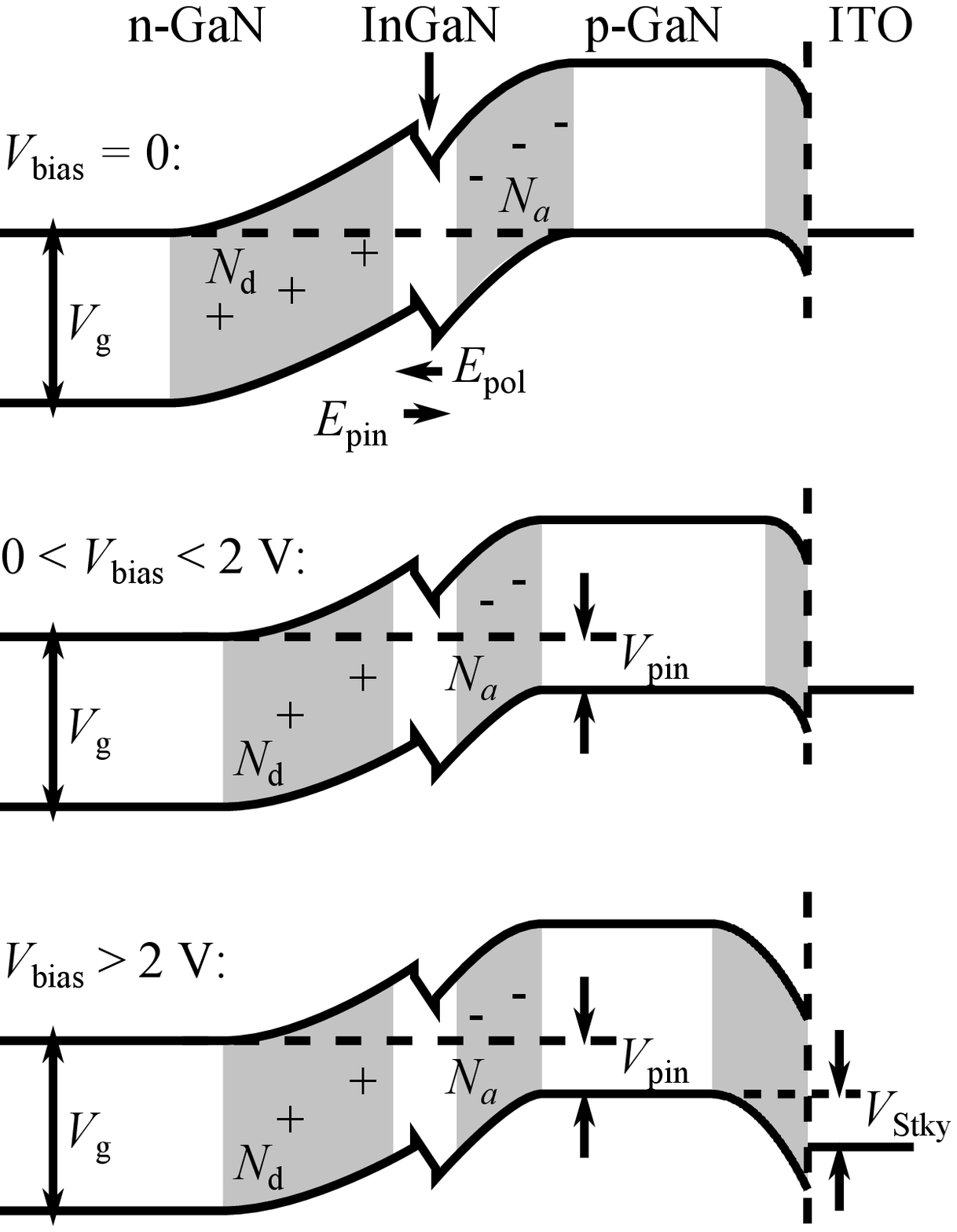}
\caption{\label{fig:theory} Schematic energy band diagrams of the nanopillar at $V_{bias}=0$~V (upper panel), $0 < V_{bias} < 2$~V (middle panel) and $V_{bias}>2$~V (lower panel).}
\end{figure}

The redshift at $V_{bias}<2$~V is attributed to the decrease of the depletion width in the pin junction as illustrated in Fig.~\ref{fig:theory}. The PL energy is influenced by both the built-in polarization field $E_{pol}$ and the external electric field $E_{pin}$ due to the donor and acceptor depletion in the n-GaN and p-GaN regions, respectively. $E_{pol}$ is determined by the spontaneous and piezoelectric polarizations in the InGaN region \cite{Bernardini2000} and is independent of $V_{bias}$. $E_{pin}$ is proportional to the depletion width in the n-GaN and p-GaN regions \cite{sze2006physics} and therefore decreases as the voltage drop across the junction, $V_{pin}$, increases. At a small bias ($V_{bias}<2$~V), $V_{bias}\approx V_{pin}$. As a result, the increase of $V_{bias}$ corresponds to a rapid shrinkage of the depletion width, leading to a decrease in $|E_{pin}|$. Due to the quantum confined Stark effect \cite{Miller1984, Chichibu1999}, a stronger total electric field $E_t = E_{pol} + E_{pin}$ leads to a lower PL energy. Since $E_{pol}$ and $E_{pin}$ have opposite directions, the redshift observed in Fig.~\ref{fig:pl} suggests that $|E_{pin}| < |E_{pol}|$. The much slower redshift at $V_{bias}>2$~V is attributed to the dominance of the p-contact resistance as explained earlier. Hence, the change of $V_{pin}$ versus $V_{bias}$ becomes negligible and $E_{pin}$ stops decreasing, leading the near absence of further PL energy shift.


\label{Conclusion}

In summary, we demonstrated electrically pumped, site- and structure-controlled QD single-photon-emitting diodes. The QDs were fabricated using a scalable fabrication process starting from a standard blue-green InGaN LED structure. The high degree of control over the QD site and dimension as well as compact device footprint make the proposed SPS suitable for on-chip integration with other electrical and optical components. Further improvements on the contact resistance, optimized doping profile to minimize background emission, and the use of AlGaN quantum barrier are needed for lower $g^{(2)}$ and high-temperature operation \cite{Deshpande2013, Holmes2014}.

\label{Acknowledgement}
We acknowledge financial supports from the National Science Foundation (NSF) under Awards DMR 1409529 for work related to materials properties and device design and DMR 1120923 (MRSEC) for work related to the measurements.

%

\end{document}